\documentstyle[preprint,aps]{revtex}
\def\lsim{\mathrel{\rlap{\lower4pt\hbox{\hskip1pt$\sim$}}
    \raise1pt\hbox{$<$}}}         
\def\gsim{\mathrel{\rlap{\lower4pt\hbox{\hskip1pt$\sim$}}
    \raise1pt\hbox{$>$}}}         

\def\overleftrightarrow#1{\vbox{\ialign{##\crcr
    $\leftrightarrow$\crcr
    \noalign{\kern 1pt\nointerlineskip}
    $\hfil\displaystyle{#1}\hfil$\crcr}}}
\def\bea{\begin{eqnarray}}
\def\beann{\begin{eqnarray*}}
\def\beq{\begin{equation}}
\def\eea{\end{eqnarray}}
\def\eeann{\end{eqnarray*}}
\def\eeq{\end{equation}}
\def\nn{\nonumber}
\font\tenbifull=cmmib10 
\font\tenbimed=cmmib10 scaled 800
\font\tenbismall=cmmib10 scaled 666
\def\boldsigma{\fam=9{\mathchar"711B } }
\def\boldtau{\fam=9{\mathchar"711C } }
\newcommand{\bcdot}{{\bf\cdot}}

\def\sig{\sigma}
\def\al{\alpha}

\def\calo{{\bf\cal O}}
\def\bp{{\bf p}}
\def\bP{{\bf P}}

\begin{document}
\draft
\preprint{\parbox[t]{85mm}{Preprint number: \parbox[t]{48mm}
{IFT-P.054/95}}}
\title{Second quantization approach to composite hadron
interactions in quark models}
\author{D. Hadjimichef\footnotemark[1], G. Krein\footnotemark[1],
S. Szpigel\footnotemark[2], and J.S. da Veiga\footnotemark[1]}
\address{\footnotemark[1]Instituto de F\'{\i}sica Te\'orica -
Universidade Estadual Paulista\\
Rua Pamplona 145, 01405-900 S\~ao Paulo-SP, Brazil\\
\footnotemark[2] Instituto de F\'{\i}sica - Universidade de S\~ao Paulo \\
Caixa Postal 20516, 01498-900 S\~ao Paulo-SP, Brazil
}
%

\maketitle

\begin{abstract}
Starting from the Fock space representation of hadron bound states in
a quark model, a change of representation is implemented by a unitary
transformation such that the composite hadrons are redescribed by
elementary-particle field operators. Application of the unitary
transformation to the microscopic quark Hamiltonian gives rise to
effective hadron-hadron, hadron-quark, and quark-quark Hamiltonians.
An effective baryon Hamiltonian is derived using a simple quark model.
The baryon Hamiltonian is free of the post-prior discrepancy which
usually plagues composite-particle effective interactions.
\end{abstract}
%
%
\newpage

In the last 20 years the studies of the hadron-hadron interaction
using quark models have been performed mainly by means of the
traditional cluster techniques such as adiabatic methods, resonating
group (RGM) or generator coordinate methods, and variational
techniques~\cite{refer}. In this paper we consider a different
approach, which was developed independently by Girardeau~\cite{girar1}
and Vorob'ev and Khomkin~\cite{russ} to deal with problems in atomic
physics where the internal degrees of freedom of atoms cannot validly
be neglected. The method employs a second quantization formalism and
shares some characteristics with Weinberg's~\cite{weinb} quasi-particle
approach and also with the ``quark Born diagram" (QBD) approach recently
introduced by Barnes and collaborators~\cite{QBD}~\cite{QBD_BB}.
Girardeau coined the name Fock-Tani (FT) representation for the
formalism.

The use of a second quantization formalism offers several advantages
over a first quantization one. Such advantages include the use of the
known field theoretic techniques such as Feynman diagrams and
Green's functions which have proven to greatly simplify the discussion
of many-body systems in different areas of physics. Since in a hadronic
collision both the constituents of the hadrons and the hadrons
themselves can participate in the intermediate states,
one expects simplifications by describing the hadrons participating
in the process in terms of macroscopic hadron field operators,
instead of the microscopic constituent ones. The problem that arises is
that composite hadron field operators in general do not satisfy canonical
(anti)commutation relations and, therefore, the traditional field theoretic
techniques cannot be directly applied. However, in the FT formalism one
recovers the possibility of using these techniques by a change of
representation
in which the composite operators are redescribed by elementary-particle
operators
satisfying canonical (anti)commutation relations. In some sense this realizes
the dual quark-hadron description of hadronic processes.

In this letter we present the extension of the original atomic physics
formalism to a general class of quark models in which the baryons are
described as bound states of three constituent quarks. In Fock space ($\cal F$)
a
one-baryon state of c.m. momentum $\bf P$, internal energy $\epsilon$,
spin projection $M_S$, and isospin projection $M_T$ is denoted by
$|\alpha \rangle \equiv |{\bf P}, \epsilon, M_S, M_T\rangle=
B_{\alpha}^{\dag}|0\rangle$, where $B_{\alpha}^{\dag}$ is the baryon
creation operator:
\beq
B_{\alpha}^{\dag} = \frac{1}{\sqrt{3!}}\,
\Phi_{\alpha}^{\mu_1\mu_2\mu_3}q_{\mu_1}^{\dag}q_{\mu_2}^{\dag}
q_{\mu_3}^{\dag},
\label{2}
\eeq
and $|0\rangle$ is the vacuum state (no quarks). A summation over repeated
indices is implied. The indices $\mu_i, \nu_i,\cdots $ denote the spatial,
spin-flavor, and color coordinates of the i-th quark.
$\Phi_{\alpha}^{\mu_1\mu_2\mu_3}$ is the baryon wave
function, which is antisymmetric in the quark indices and
orthonormalized. While the quark operators $q^{\dag}_{\mu}$ and
$q_{\mu}$ satisfy the canonical anticommutation relations, the baryon
operators satisfy the following {\em noncanonical} anticommutation
relations:
\beq
\{B_{\alpha},B_{\beta}^{\dag}\}=\delta_{\alpha\beta}-
\Delta_{\alpha\beta},\hspace{1.0cm}
\{B_{\alpha},B_{\beta}\}=\{B^{\dag}_{\alpha},
B^{\dag}_{\beta}\}=0,
\label{Bcomm}
\eeq
where
\beq
\Delta_{\alpha\beta}=3\,\Phi_{\alpha}^{*\mu_1\mu_2\mu_3}
\Phi_{\beta}^{\mu_1\mu_2\nu_3}q_{\nu_3}^{\dag}q_{\mu_3}
-\frac{3}{2}\,\Phi_{\alpha}^{*\mu_1\mu_2\mu_3}
\Phi_{\beta}^{\mu_1\nu_2\nu_3}q_{\nu_3}^{\dag}q_{\nu_2}^{\dag}
q_{\mu_2}q_{\mu_3}.
\label{Delta}
\eeq
The presence of the term $\Delta_{\alpha\beta}$ is the physical manifestation
of
the internal structure of the baryons. Because of this term, the usual field
theoretic techniques, such as Wick's theorem and Greens's functions, cannot
be directly applied to the baryon operators $B$ and $B^{\dag}$. To circumvent
these problems with the noncanonical nature of the baryon operators, a change
of representation is implemented by means of a unitary transformation $U$. Of
course, the effects induced by the term $\Delta$ will appear in the effective
Hamiltonians describing the interactions among composites and constituents.
The transformation $U$ is such that a {\em single} real-baryon state
$|\alpha\!>=B^{\dag}_{\alpha}|0\!>$ is transformed into a {\em single}
ideal-baryon state $|\alpha)=b^{\dag}_{\alpha}|0)\equiv U^{-1}|\alpha\!>$.
The operator $U$ is of the general form~\cite{{girar1},{russ}}:
\beq
U=\exp\left( -\pi/2\,F \right), \hspace{1.0cm}
F=O^{\dag}_{\alpha}b_{\alpha}-b^{\dag}_{\alpha}O_{\alpha}.
\label{uandf}
\eeq
The $b^{\dag}_{\alpha}$ and  $b_{\alpha}$ are the ideal-baryon
creation and annihilation operators and $O^{\dag}_{\alpha}$ and
$O_{\alpha}$ are functionals of the $B^{\dag}_{\alpha}$,
$B_{\alpha}$ and $\Delta_{\alpha\beta}$ operators. The $b$'s and
$O$'s satisfy canonical anticommutation relations:
\bea
&&\{b_{\alpha}, b^{\dag}_{\beta}\}=\{O_{\alpha}, O^{\dag}_{\beta}\}=
\delta_{\alpha\beta},\nn\\
&&\{b_{\alpha}, b_{\beta}\}=\{b^{\dag}_{\alpha}, b^{\dag}_{\beta}\}=
\{O_{\alpha}, O_{\beta}\}=\{O^{\dag}_{\alpha}, O^{\dag}_{\beta}\}=0,
\eea
and, by definition, the $b$ and $b^{\dag}$ anticommute with the quark
operators. $U$ acts
on an enlarged Fock space ${\cal I}$, which is the graded direct product
of the original Fock space ${\cal F}$ and an ideal state space ${\cal B}$
(the space spanned by the ideal baryons). The vacuum state of space $\cal I$,
$|0)$, is the direct product of the vacua of $\cal F$ and $\cal B$.
In ${\cal I}$ the physical states, $|\psi\!>$, constitute a subspace
isomorphic to the original Fock space ${\cal F}$, and satisfy the
constraint $b_{\alpha}|\psi\!>=0$. The unitary transformation acting
on the physical states gives rise to the subspace ${\cal F}_{\rm FT}=
U^{-1}{\cal I}_0$, where the physical states are required to satisfy
the transformed constraint: $U^{-1}b_{\alpha}|\psi)=0$, where $|\psi)
=U^{-1}|\psi\!>$. In ${\cal F}_{\rm FT}$ all operators satisfy
canonical (anti)commutation relations and, therefore, the traditional
field theoretical methods can be employed. In summary, the effect of the
transformation on the real baryon states is to redescribe these by ideal
ones, and since one simultaneously transforms the operators written
in terms of the microscopic quark operators, such as the Hamiltonian,
electroweak currents, etc, expectation values and matrix elements are
preserved because the transformation is unitary. The advantage, as said above,
is that all operators in the new representation are canonical, and the role
played by the bound states in the processes is made explicit. A more detailed
discussion of these and other formal issues can be found in
Refs.~\cite{{girar},{jmp},{straton}}.

$O_{\alpha}$ is constructed by an iterative procedure as a power series
in the baryon wave functions $\Phi$: $O_{\alpha}=\sum_n
O^{(n)}_{\alpha}$, where $n$ identifies the power of $\Phi$ in the
expansion. The expansion starts at zeroth-order with $O^{(0)}_{\alpha}=
B_{\alpha}$. The construction of the higher order terms
$O^{(n)}_{\alpha}$, $n \geq 1$, involves addition of a series of
counterterms such that anticommutation relations of $O$ and $O^{\dag}$
are satisfied order by order. Since $\{O^{(0)}_{\alpha},
O^{(0)\dag}_{\beta}\}=\delta_{\alpha\beta} - \Delta_{\alpha\beta}$,
and $\Delta_{\alpha\beta}$ is of second order [see Eq.~(\ref{Delta})],
one has that $O^{(1)}_{\alpha}=0$, and the next nonzero term is then
of order $n=2$. It is not difficult to show that the second order
counterterm that has to be added to $O^{(0)}_{\alpha}$ to cancel the
$\Delta_{\alpha\beta}$ in $\{O^{(0)}_{\alpha},O^{(0)\dag}_{\beta}\}$
is equal to $1/2\Delta_{\alpha\beta}B_{\beta}$. Then, up to $n=2$,
$O_{\alpha}= B_{\alpha}+1/2\Delta_{\alpha\beta}B_{\beta}$ and one
obtains $\{O_{\alpha},O^{\dag}_{\beta}\}=\delta_{\alpha\beta}-
1/2[\Delta_{\alpha\gamma},B_{\beta}]B_{\gamma}-1/2B^{\dag}[B_{\alpha},
\Delta_{\gamma\beta}]=\delta_{\alpha\beta}+{\cal O}(\Phi^3)$. A third
order counterterm has to be added such that the ${\cal O}(\Phi^3)$
piece cancels, and so on to higher orders. However, for our purposes
here one needs $O_{\alpha}$ up to $n=3$ only:
\bea
O_{\alpha}= B_{\alpha} +\frac{1}{2}\Delta _{\alpha\beta}B_{\beta} -
\frac{1}{2}B^{\dag}_{\beta}[\Delta_{\beta\gamma},B_{\alpha}]
B_{\gamma}\;.
\label{O3}
\eea

When the unitary transformation is applied to the microscopic quark
Hamiltonian, one obtains effective Hamiltonians which describe all
possible processes involving hadrons and their constituents. We
discuss this for a microscopic quark Hamiltonian in which quarks interact
by two-body forces. Such a Hamiltonian can always be written as:
\bea
H = T(\mu) q_\mu ^{\dag }q_\mu +\frac 12 V(\mu \nu; \sig\rho) \;
q_\mu ^{\dag }q_\nu ^{\dag }q_\rho q_\sig,
\label{Hmic}
\eea
where $T$ is the kinetic energy and $V_{qq}$ is the quark-quark
interaction. In free space, $\Phi$ satisfies the equation of motion:
\beq
H(\mu \nu;  \sigma \rho )\Phi_\alpha^{ \sigma \rho \lambda}
=3\left[\;\delta_{[\mu] \sig }\delta_{\nu\rho}\; T ([\mu])
+V_{qq}(\mu \nu;  \sig\rho)\;\right]\Phi_\alpha^{\sigma\rho\lambda}
=E_{[\alpha]} \Phi_{[\alpha]}^{\mu\nu\lambda},
\label{Hmat}
\eeq
where we are using the convention that there is no sum over repeated
indices inside square brackets, and $E_{\alpha}$ is the total energy
(center-of-mass energy plus internal energy) of the baryon.
The transformation of the Hamiltonian is made by transforming
initially the quark operators $q$ and $q^{\dag}$. Since the $O$
operators are given by a power series, the transformed quark operators
are also obtained as a power series:
$U^{-1}q_{\mu}U=\sum_n q^{(n)}_{\mu}$. The $q^{(n)}_{\mu}$ can be
obtained by expanding the exponential in Eq.~(\ref{uandf}) to the
desired order or, equivalently, by means of the ``equation of motion"
technique~\cite{girar}. Up to third order, one obtains:
\bea
q^{(1)}_{\mu} &=& -\,\sqrt{\frac{3}{2}}
\Phi^{\mu\mu_{2}\mu_{3}}_{\alpha}q^{\dag }_{\mu_{2}} q^{\dag}_{\mu_{3}}
\left( b_{\alpha} + B_{\beta} \right),\nn\\
q^{(2)}_{\mu} &=& \frac{3}{2}\Phi^{\ast\mu_{1}\mu_{2}\mu_{3}}_{\alpha}\left[
\Phi^{\mu_{1}\mu_{2}\mu}_{\beta}\left(b^{\dag}_{\alpha}q_{\mu_{3}}
b_{\beta}+
B^{\dag}_{\alpha}q_{\mu_{3}}B_{\beta}+2B^{\dag}_{\alpha}q_{\mu_{3}}
b_{\beta}\right)\right.\nn\\
&-&\left.
\Phi^{\mu_{1}\nu_{2}\mu}_{\beta} \left( b^{\dag}_{\alpha}
q^{\dag}_{\nu_{2}}q_{\mu_{2}}q_{\mu_{3}} b_{\beta}+
B_\alpha^{\dagger }q_{\nu_2}^{\dagger}q_{\mu_2}q_{\mu_3}
B_\beta + 2B_\alpha^{\dagger }q_{\nu_2}^{\dagger}q_{\mu_2}
q_{\mu_3}b_\beta \right)\right],\nn\\
\label{q2}
q_{\mu}^{(3)} &=& -\sqrt{\frac{27}{8}}\Phi_\alpha^{\mu_1\mu_2\mu_3}\left[
\frac{1}{3} q_{\mu_2}^{\dag}q_{\mu_3}^{\dag}
\Delta_{\alpha\beta}\left(2B_\beta+b_\beta\right)
-\Phi_\beta^{*\nu_1\nu_2\nu_3}\Phi_\gamma^{\mu\mu_2\nu_3}
\left(B_\alpha^{\dagger }+b^{\dag}_{\alpha}\right)\right.\nn\\
&\times& B_\beta^{\dag}q_{\mu_1}q_{\mu_3}q_{\nu_2}q_{\nu_1}b_\gamma
+\left(2\Phi_\beta ^{\mu_1\mu_2\nu_3}
\Phi_\gamma ^{\mu \nu_2\mu_3}-
\Phi_\beta ^{\mu_1\nu_2\nu_3}
\Phi_\gamma^{\mu\mu_2\mu_3}\right)  \nn\\
&\times& \left(b_\alpha^{\dag}q_{\nu_2}^{\dag}q_{\nu_3}^{\dag}b_\beta
b_\gamma
+B_\alpha^{\dag}q_{\nu_2}^{\dag}q_{\nu_3}^{\dag}B_\beta B_\gamma
+B_\alpha^{\dag}q_{\nu_2}^{\dag}q_{\nu_3}^{\dag}B_\beta b_\gamma
+ b_\alpha^{\dag}q_{\nu_2}^{\dag}q_{\nu_3}^{\dag}B_\beta b_\gamma
\right)\nn\\
&-& 2\, \Phi_\beta^{\mu_1\nu_2\nu_3}\Phi_\gamma^{\mu\tau_2\mu_3}\left(
b_\alpha^{\dag}q_{\tau_2}^{\dag}q_{\nu_3}^{\dag}q_{\nu_2}^{\dag}
q_{\mu_2}b_\beta b_\gamma
+ B_\alpha^{\dag}q_{\tau_2}^{\dag}q_{\nu_3}^{\dag}q_{\nu_2}^{\dag}
q_{\mu_2}B_\beta B_\gamma\right.\nn\\
&+& \left.\left. B_\alpha^{\dag}q_{\tau_2}^{\dag}q_{\nu_3}^{\dag}
q_{\nu_2}^{\dag}q_{\mu_2}B_{\beta} b_{\gamma} \right)\right].
\label{q3}
\eea

Substituting these in Eq.~(\ref{Hmic}), one obtains that the general
structure of the transformed Hamiltonian is:
\beq
U^{-1}HU=H_{q}+ H_{b} +H_{bq},
\label{Htrans}
\eeq
where the subindices identify the operator content of each term. $H_q$
describes true quark-quark scattering only, i.e., it is unable
to bind the quarks into the bound state baryons; the bound states appear
in $H_b$. This feature leads~\cite{qchem} to the same effect of curing the
bound state divergences of the Born series as in Weinberg's quasi-particle
method~\cite{weinb}. $H_{bq}$ describes quark-baryon processes as baryon
breakup into three quarks and three quarks recombining into a baryon. In models
where quarks are confined, these terms contribute to free-space baryon-baryon
processes as intermediate states only. However, in high temperature
and/or density systems hadrons and quarks can coexist and the breakup
and recombination processes can play important role. Work is in
progress where the explicit form and applications involving the terms
$H_q$ and $H_{qb}$ are discussed. Next we discuss the baryon Hamiltonian $H_b$.

Using the transformed quark operators given in
Eqs.~(\ref{q3}), one obtains the effective baryon
Hamiltonian:
\bea
H_b=\Phi^{*\mu\nu\lambda}_{\alpha}H(\mu\nu;\sigma\rho)
\Phi^{\sigma\rho\lambda}_{\beta}\;b^{\dag}_{\alpha}b_{\beta}+
\frac{1}{2}\;V_{bb}(\al\beta;\delta\gamma)\;b^{\dag}_{\al}
b^{\dag}_{\beta}b_{\gamma}b_{\delta}\;,
\label{Hb}
\eea
where $V_{bb} \equiv V^{dir}+V^{exc}+V^{int}$ is an effective baryon-baryon
potential,
which we divide for later convenience into direct, exchange, and intra-exchange
parts:
\bea
&&V^{dir}_{bb}(\alpha\beta;\delta\gamma) = 9 \; V_{qq}(\mu \nu; \sigma \rho)
\Phi^{\ast\mu\mu_{2}\mu_{3}}_{\alpha}\Phi^{\ast\nu\nu_{2}\nu_{3}}_{\beta}
\Phi^{\rho\nu_{2}\nu_{3}}_{\gamma}\Phi^{\sigma\mu_{2}\mu_{3}}_{\delta},\\
\label{Vdir}
&&V^{exc}_{bb}(\alpha\beta;\delta\gamma) =
9 \; V_{qq}(\mu \nu; \sigma \rho)\left\{2\,\Phi^{\ast\mu\nu\mu_{3}}_{\alpha}
\Phi^{\ast\nu_{1}\nu_2\nu_{3}}_{\beta} \Phi^{\rho\nu_2\nu_3}_{\gamma}
\Phi^{\nu_1\sigma\mu_3}_{\delta}-\Phi^{\ast\mu\mu_{2}\mu_{3}}_{\alpha}\right.\nn\\
&&\;\;\;\;\times \left.\left[
\Phi^{\ast\nu\nu_{2}\nu_{3}}_{\beta}\left(\Phi^{\sigma\nu_{2}\nu_{3}}_{\gamma}
\Phi^{\rho\mu_{2}\mu_{3}}_{\delta}+4\,(\sigma\leftrightarrow\rho)\right)
- 2 \,\Phi^{\ast\mu_{1}\nu\nu_{3}}_{\beta} \Phi^{\sigma\rho\nu_{3}}_{\gamma}
\Phi^{\mu_{1}\mu_2\mu_3}\right]\right\},\\
\label{Vexch}
&&V^{int}_{bb}(\alpha\beta; \delta\gamma)=-3 \; H(\mu \nu ;\sigma\rho)
\left(\Phi_{\alpha}^{*\mu\nu\mu_3}\Phi_{\beta}^{*\nu_1\nu_2\nu_3}
\Phi_{\gamma}^{\nu_1\nu_2\mu_3}\Phi_{\delta}^{\sigma\rho\nu_3}\right. \nn\\
&&\;\;\;\;+\left. \Phi_{\alpha}^{*\mu\mu_2\nu}\Phi_{\beta}^{*\nu_1\nu_2\nu_3}
\Phi_{\gamma}^{\nu_1\nu_2\rho}\Phi_{\delta}^{\sigma\mu_2\nu_3}
+\Phi_{\alpha}^{*\mu\mu_2\nu_3}\Phi_{\beta}^{*\nu_1\nu_2\nu}
\Phi_{\gamma}^{\nu_1\nu_2\nu_3}\Phi_{\delta}^{\sigma\mu_2\rho}\right).
\label{Vintr}
\eea
For reasons that will become clear in the discussion below, in
Eqs.~(\ref{Hb},\ref{Vintr}) we are not supposing that the $\Phi$'s
are eigenstates of the microscopic quark Hamiltonian, as in Eq.~(\ref{Hmat}).

The higher order terms which are neglected in Eq.~(\ref{O3}) give rise
to effective many-baryon (higher than two-baryon) forces, and also
introduce orthogonality corrections. These orthogonality corrections
are similar to the ``wave function renormalization" of RGM
calculations~\cite{refer} and have the effect of weakening the
``intra-exchange" interactions, i.e., interactions where the
microscopic quark-quark interaction occurs within a single
composite~\cite{{girar},{jmp},{straton}}. When the $\Phi$'s are
eigenstates of the microscopic quark Hamiltonian, the lowest order
orthogonalization cancels the term $V^{int}_{bb}$. The weakening of
the intra-exchange terms is the major effect of the orthogonalization;
higher order corrections start at ${\cal O}(\Phi^8)$ and vanish
asymptotically with increasing baryon energy.

Eqs.~(\ref{O3},\ref{q3},\ref{Hb}-\ref{Vintr}) embody the
main results of the present paper; these are valid for any quark model
which describes the nucleon as a three quark system, and quarks interacting
by two-body forces. The $\Phi$'s are not restricted to ground states, they
describe the entire baryon spectrum derived from the microscopic quark
Hamiltonian.

An important feature of the effective baryon Hamiltonian is that
it gives rise to scattering amplitudes that are symmetric under
exchange of initial and final channels. The lack of this symmetry is
known as the post-prior discrepancy~\cite{post-prior}. The discrepancy
is catastrophic for processes with initial and final states with
different masses. It is not difficult to convince oneself of this
property by exchanging $\alpha$ and $\beta$  with $\gamma$ and
$\delta$ in the expressions above. In the example that follows
the symmetry is evident since the effective potential is symmetric
under exchange of initial and final momenta (${\bf p} \leftrightarrow
{\bf p'}$) [see Eq.~(\ref{gra5}) below].

To finalize, we present the result of a derivation~\cite{jmp} of an
effective nucleon-nucleon interaction following the model of Barnes
et al.~\cite{QBD_BB}. These authors restrict the $\Phi$'s to
nonrelativistic s-wave gaussians and use for the microscopic
quark-quark interaction the spin-spin part derived from the
nonrelativistic reduction of the one-gluon exchange. In second-quantized
notation this interaction can be written as:
\bea
V_{qq}&=&\frac{1}{2} \frac{8\pi\alpha_s}{3m_q^2}
\int \frac{d^3{\bp}_1 d^3{\bp}_2 d^3{\bp}_3d^3{\bp}_4}
{(2\pi)^3} \; \delta({\bp}_1 + {\bp}_2 - {\bp}_3 - {\bp}_4)\,
\delta_{t_1 t_3}\delta_{t_2 t_4} \;\nn\\
&\times & \frac{1}{4} \lambda^a_{c_1 c_3}\lambda^a_{c_2 c_4} \,
\frac{1}{4}{\boldsigma}_{s_1 s_3}\cdot{\boldsigma}_{s_2 s_4}\,
q^{c_1\dag}_{s_1 t_1}(\bp_1) q^{c_2\dag}_{s_2 t_2}(\bp_2)
q^{c_4}_{s_4 t_4}(\bp_4) q^{c_3}_{s_3 t_3}(\bp_3),
\eea
The restriction to this part of the full quark-quark interaction Hamiltonian
which in principle should include confinement and other spin-dependent
components, is because previous calculations have shown that the
spin-spin component provides the dominant contribution to s-wave NN
scattering.

Supposing that $\Phi$ is an eigenstate of the full quark Hamiltonian,
one is left to lowest order with the $V^{exc}$ part only for the NN
effective interaction. Because the single-quark wave functions are
gaussians and the quark-quark interaction is a constant (in momentum
space), the 12-dimensional integral can be
evaluated analytically. The sum over the quark color-spin-flavor
indices can be done in closed form using the elegant technique of
Ref.~\cite{hol}. The final result for the effective NN potential is
given by:
\bea
V_{NN}&=&\frac{1}{2}\int d^3{\bP} d^3{\bP'}d^3{\bp}d^3{\bp'}\;
\delta(\bP'-\bP)<\!\lambda_1\lambda_2|V_{NN}({\boldsigma}, {\boldtau}, {\bp},
{\bp'})|\lambda_3\lambda_4\!>\nn\\
&\times & b^{\dag}_{\lambda_1}(\bp'+\bP'/2)\,b^{\dag}_{\lambda_2}(\bp'-\bP'/2)
\,b_{\lambda_4}(\bp-\bP/2)\,b_{\lambda_3}(\bp+\bP/2),
\label{VNNNL}
\eea
where $\lambda=(M_S,M_T)$ and $V_{NN}(\boldsigma, \boldtau, \bp, \bp')=
\kappa_{ss}\sum_{i=1}^{5}\, \calo_{i}(\boldsigma,\boldtau)
v_{i}({\bp}, {\bp'})$, where
$\kappa_{ss}={8\pi\alpha_s}/{3m_q^2(2\pi)^3}$,
$\calo_1=0$, and:
\bea
\calo_{2}&=&\frac{1}{12}\left[\left(1+\frac{1}{9}{\boldtau}^{1}_{N}\bcdot
{\boldtau}^{2}_{N}\right)+ \frac{1}{3}\left(1+
\frac{1}{9}{\boldtau}^{1}_{N}\bcdot
{\boldtau}^{2}_{N}\right){\boldsigma}^{1}_{N}\bcdot
{\boldsigma}^{2}_{N}\right],\nn\\
\calo_{3}&=&\frac{3}{4}\left[\left(1+\frac{1}{9}{\boldtau}^{1}_{N}\bcdot
{\boldtau}^{2}_{N}\right)- \frac{1}{27}\left(1+ \frac{25}{9}{\boldtau}^{1}_{N}
\bcdot{\boldtau}^{2}_{N}\right){\boldsigma}^{1}_{N}\bcdot{\boldsigma}^{2}_{N}\right],\nn\\
\calo_{4}&=&\calo_{5}=
\frac{1}{4}\left[\left(1-\frac{1}{9}{\boldtau}^{1}_{N}\bcdot
{\boldtau}^{2}_{N}\right)-\frac{1}{9}\left(1-
\frac{5}{9}{\boldtau}^{1}_{N}\bcdot
{\boldtau}^{2}_{N}\right){\boldsigma}^{1}_{N}\bcdot{\boldsigma}^{2}_{N}\right]
\label{gra4}
\eea
$\boldsigma_N$ and $\boldtau_N$ are nucleon spin and isospin operators, and
\bea
v_1({\bp}, {\bp}')&=& v_3({\bp},{\bp}')=\exp\left[-\frac{a^2}{3}
\left(\bp-\bp'\right)^2\right], \nn\\
v_2({\bp}, {\bp}')&=& \left(\frac{3}{4}
\right)^{\frac{3}{2}}\exp \left[ -\frac{a^2}{6} \left({\bp}^2+
{\bp'}^2\right)\right],\nn\\
v_4({\bp}, {\bp}')&=&v_5({\bp'},{\bp})=
\left(\frac{12}{11}\right)^{\frac{3}{2}}\exp \left[-\frac{2a^2}{11}
\left(\bp-\bp'\right)^{2}-\frac{a^2}{33}
\left( {\bp}^{2} + 7 {\bp'}^2\right) \right],
\label{gra5}
\eea
and $a$ is the r.m.s. radius of the nucleon.
$\calo_{1}=0$ because of color: there is no one gluon exchange
between colorless baryons. This term corresponds to the direct part
$V^{dir}$ of the baryon-baryon interaction.

{}From the expressions for the $v_i({\bp}, {\bp'})$, it is clear that
the interaction is symmetric under $\bp \leftrightarrow \bp'$ and,
therefore, free from the post-prior discrepancy. The effective
baryon-baryon interaction $V_{NN}$ is similar to the one derived
by Barnes et al.~\cite{QBD_BB} using the quark Born diagram technique.
When calculating the on-shell (${\bp}^2={\bp'}^2$) Born-order T-matrix
with the above effective interaction, we arrive at their expressions
in~\cite{QBD_BB}. Barnes et al. calculated phase shifts by
appropriately defining a local NN potential and obtained results
qualitatively similar to the RGM ones. We have also calculated
phase shifts, however, we solved numerically the Lippman-Schwinger
equation and used the full non-local NN interaction above. Using the
parameter set of Barnes et al.~\cite{QBD_BB}, $\alpha_s/m_q^2=0.6/(330)^2$
MeV$^{-2}$, $a=0.5$ fm and $m_q=330$ MeV, we obtain for the s-wave phase
shifts $^1S_0$ and $^3S_1$ the results shown in the figure below.

\vskip 0.5cm
\begin{center}
\fbox{Figure}
\end{center}
\noindent
Figure 1. S-wave phase-shifts from the $V_{NN}$ of
Eqs.~(\ref{VNNNL}-\ref{gra5}).
Solid line is for S=0, I=1 and dashed is for S=1, I=0.
Parameter set: $\alpha_s/m_q^2=0.6/(330)^2$ MeV$^{-2}$, $a=0.5$ fm and
$m_q=330$ MeV.
\vskip 0.5cm

The nature of the repulsive core of the NN interaction is clearly seen in this
figure. The magnitude of the phase shifts are quantitatively similar to the RGM
ones. Moreover, when a purely phenomenological long-range attractive
quark-quark
interaction is added to the spin-spin one-gluon exchange interaction, we can
fit
the experimental low-energy s-wave phase shifts by adjusting the parameters of
the atractive interaction. We have also checked the effect of the post-prior
discrepancy in asymmetrical systems. We found that the lack of the
initial-final
state symmetry induces large errors in scattering cross sections.

Similar techniques to the one discussed here, mainly developed in the
context of nuclear structure, have recently being adapted to the hadronic
physics problems~\cite{nucl}. However their emphasis and aim are different
from ours. In particular, references~\cite{nucl} deal with the
problem of constructing a transformation such that many real-hadron
states are mapped to many ideal-hadron states, whereas here the transformation
is constructed such that a single real-hadron state is mapped into a singe
ideal-hadron state.

In this paper we have applied the formalism to a simple
example, however, the formalism is sufficiently powerful and practical to
be used in more realistic studies of composite hadrons interactions. In
particular, the formalism should be useful for  studying hadronic interactions
using relativistic light-cone quark models. The Fock space structure
of the hadronic states in such models is similar to the ones considered in this
paper. Also, the study of the high temperature and/or density regime
of hadronic matter, where hadrons and quarks can coexist, by using
standard many-body techniques with the effective Hamiltonians derived
with the present formalism is particularly interesting.

\vspace{1.0cm}
\noindent
{\bf Acknowledgments}

The authors thank Professor Girardeau for many discussions and
clarifications on the FT formalism. The authors also thank Drs.
S.K. Adhikari, M. Betz, M.R. Robilotta and L. Tomio for discussions.
GK is grateful to K. Holinde and K. Yazaki for discussions on
cluster techniques. This work was partially supported by CAPES,
CNPq and FAPESP.


%

\end{document}